\documentclass{jltp}

\usepackage{epsfig}

\title{
Annular Long Josephson Junctions in a Magnetic Field:
Engineering and Probing the
Fluxon Interaction Potential}

\author{A.~Wallraff,
Yu.~Koval,
M.~Levitchev$^{*}$,
M.~V.~Fistul,\\
and A.~V. Ustinov}

\address{Physikalisches Institut III, Universit{\"a}t 
Erlangen-N{\"u}rnberg,\\ D-91058 Erlangen, Germany\\
$^{*}$ISI, Forschungszentrum J{\"u}lich GmbH, D-52425 
J{\"u}lich,\ Germany}

\runninghead{A.~Wallraff \textit{et al.}}
{Annular Long Josephson Junctions in a Magnetic Field}

\begin{document}

\maketitle

\begin{abstract}
The interaction of a Josephson fluxon with an external magnetic
field-induced potential in a long Josephson junction is investigated
experimentally.  The thermal activation of the fluxon from a potential
well is observed and experiments probing its predicted quantum
properties are discussed.  A method for engineering a magnetic
double-well potential for a fluxon is proposed and the use of the
coupled fluxon states for quantum computation is suggested.
    
PACS numbers: 
  74.50.+r,  %Josephson effects
  05.45.Yv,  %Solitons
  85.25.Cp,  %Josephson devices
  03.67.Lx.  %Quantum Computation
\end{abstract}

\section{INTRODUCTION}
Long Josephson junctions are known as the most clean systems to
experimentally study basic properties of Josephson vortices, often
also called fluxons.\cite{Ustinov98}  The interaction of a Josephson
vortex with a potential induced by an externally applied magnetic
field has been discussed in literature in various
contexts.\cite{Gronbech91b,Ustinov96c}  Recently, first measurements
reporting the thermally activated escape of the vortex from a magnetic
field-induced potential well at $4.2 \, \rm{K}$ have been
reported.\cite{Fistul00}  Experiments probing the quantum properties
of a Josephson vortex trapped in a field-engineered potential are
currently in progress.

Our interest in this subject is related to a possibility of
using macroscopic quantum states of Josephson vortices for quantum
computation\cite{Lloyd93,DiVincenzo95}.  We suggest that two distinct
states of a fluxon trapped in a field-controlled double-well potential
inside a narrow long junction can be used for designing a qubit, the
basic unit of information in a quantum computer. We propose that
by varying the external magnetic field and junction shape one is able
to form an arbitrary-shaped potential for a fluxon in a junction; the
amplitude of this potential can be varied in experiment by tuning the
magnetic field.

In this paper we present our experimental results on the thermal
escape of a fluxon from a potential well measured at different
temperatures.  We also discuss the quantum tunneling of a fluxon and
some preliminary ideas on the engineering of a double-well potential
for a fluxon.

\subsection{A Fluxon in an External Potential}
We consider a long Josephson junction of arbitrary shape and constant
width $w$.  If $w$ is smaller than the Josephson penetration depth
$\lambda_{J}$ the junction can be considered as quasi-one dimensional. 
Its electrodynamics is described by a wave equation for the
superconducting phase difference $\phi(\tilde{x},\tilde{t})$ across
the junction
\begin{equation}
	\phi_{\tilde{x}\tilde{x}} - \phi_{\tilde{t}\tilde{t}} - \sin\phi = 
	\alpha \phi_{\tilde{t}} -
	\gamma + 
	f(\tilde{x},\tilde{t}).
	\label{eq:sineGordon}
\end{equation}
In this representation the time and the space coordinates are
normalized with respect to the inverse plasma frequency $t = \tilde{t}
\omega_{p}^{-1}$ and the Josephson length $x = \tilde{x}\lambda_{J}$,
respectively.\cite{Ustinov98}  In these units, the characteristic
phase velocity of electromagnetic waves in the junction is unity and
is usually called the Swihart velocity.\cite{Ustinov98}  The left hand
side of Eq.~(\ref{eq:sineGordon}) is the sine--Gordon equation (SGE). 
The right hand side of Eq.~(\ref{eq:sineGordon}) contains the
perturbation terms.  The term $\gamma = I/I_{c}$ is the bias current
normalized to the critical current, the damping term due to the
quasiparticle resistance $\rho$ is proportional to $\alpha$.  The term
$f(\tilde{x},\tilde{t})$ describes the interaction of the Josephson
vortex with a spatial inhomogeneity region in the junction.  In
absence of perturbations the Josephson vortex is described by the
exact soliton solution of the left hand side of Eq.~(\ref{eq:sineGordon})
\begin{equation}
	\phi(\tilde{x},\tilde{t}) = 4 \arctan 
	\left[\exp\left(\frac{\tilde{x} - q(\tilde{t})}
	{\sqrt{1-u^{2}}} \right)\right],
	\label{eq:soliton}
\end{equation}
where $q(\tilde{t})$ is the coordinate of the vortex and $u$ is its
velocity.

Using perturbation theory to solve the perturbed
SGE\cite{McLaughlin78} (\ref{eq:sineGordon}) one can write down an
equation of motion for the center-of-mass coordinate $q(\tilde{t})$ of
the vortex in the non-relativistic limit $(u \ll 1)$
\begin{equation}
    m_{f} \ddot{q} + \alpha \dot{q} + 
    \frac{\partial U(q)}{\partial q}=0.
	\label{eq:ofMotion}
\end{equation}
Because of the magnetic flux quantization, the trapped vortex behaves
as a topologically stable, particle-like object of mass $m_{f}$ moving
under the action of external forces.  The force acting on the vortex
can be expressed in terms of the potential $U$.  The bias current
through the junction acts as a driving force (of the Lorentz type) on
the vortex.  The pinning force may have different physical origins.  A
pinning center for a Josephson vortex in a long junction can be
realized by introducing a micro-short\cite{Galpern84} or a
micro-resistor\cite{Malomed90} in the junction barrier.  Pinning may
also occur due to the interaction of the vortex with the junction
leads\cite{Muenter98} or interaction with parasitic magnetic flux
trapped in the superconducting films.  Alternatively, a vortex can be
pinned by its magnetic dipole interaction with an external magnetic
field.\cite{Gronbech91b,Ustinov96c}  In the following, we consider the
latter pinning induced by an external magnetic field, because this
method allows to \emph{control} the pinning potential for the vortex
during the experiment.

\subsection{Magnetic Field Interaction}
The interaction of a Josephson vortex with an external magnetic field
applied in the plane of an annular junction can be taken into account by
the perturbation term\cite{Gronbech91b,Martucciello96b}
\begin{equation}
	f(\tilde{x}) = h \Delta \sin \left(\frac{2 \pi \tilde{x}}{l} \right)
	\label{eq:magneticInteraction}
\end{equation} 
in Eq.~(\ref{eq:sineGordon}), where $l = 2 \pi r/\lambda_{J}$ is the
normalized circumference of the Josephson junction, $h$ is the
magnetic field normalized to $H_{0} = \Phi_{0}/({2\pi \mu_{0} r d})$,
where $d$ is the magnetic thickness of the junction, and $\Delta$ is a
geometry dependent magnetic field coupling
coefficient.\cite{Martucciello96b}

Equivalently, in Eq.~(\ref{eq:ofMotion}) the fluxon potential $U$ depends on 
the bias current and the magnetic field as given by
\begin{equation}
	U = - \int_{0}^{l} \left( \gamma \phi +
	\Delta \phi_{\tilde{x}}  \vec{n} \cdot \vec{h}  \right) \, d\tilde{x},
	\label{eq:magFieldPotential}
\end{equation}
where $\vec{n}$ is the vector normal to the junction and lying in the
junction plane, and $\vec{h}$ is the normalized field vector.  In the
limit of local interaction ($\lambda_{J}\rightarrow 0$), the magnetic
field part of the potential can be expressed as $U = - \mu_{0}
\vec{\mu}\cdot \vec{H}$, where $\vec{\mu}$ is the magnetic moment of
the vortex normal to the boundary of the junction.

\begin{figure}[tb]
\centering
\epsfig{file=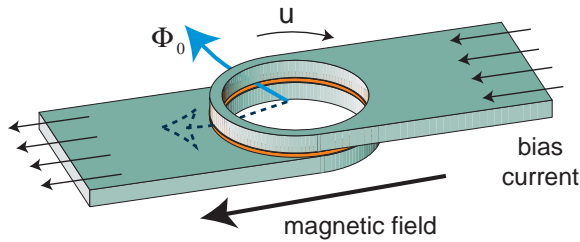,width=0.6\columnwidth}
\caption{Annular Josephson junction with a trapped Josephson vortex
(fluxon). Under the action of a bias current $I > I_{c}$ the vortex moves with
a velocity $u$ generating a voltage $V \propto u$ across the junction 
electrodes. In the static state the vortex is pinned in the direction 
parallel to the external magnetic field as indicated by the dashed arrow.}
\label{fig:exp}
\end{figure}

In the annular Josephson junction shown in Fig.~\ref{fig:exp}, the
fluxon has the lowest energy in the location where $\vec{\mu}$ is
directed along $\vec{H}$ and the highest energy in the diametrically
opposite location where $\vec{\mu}$ is opposite to $\vec{H}$.  The
pinning potential for the vortex becomes $U(q)\sim - h \cos(2\pi
q/\ell)$.  Subject to a driving force mediated by the bias current
$I$, the potential is tilted proportionally to $I$, effectively
lowering the potential barrier for the fluxon.  Thus, the fluxon
dynamics in an annular junction placed in a magnetic field is
analogous to the motion of a particle in a tilted wash-board potential
and can be mapped to the dynamics of the phase in a small
Josephson junction.\cite{Martinis87}

\section{THERMAL ACTIVATION}
At high enough temperatures, a Josephson vortex trapped in a
potential well can escape from the well by a thermally activated
process.  The thermal activation rate can be
expressed as\cite{Kramers40,Haenggi}
\begin{equation}
	\Gamma_{\rm{th}}=\frac{\omega_0 \omega_{p}}{2\pi}
	\exp\left(\frac{- U_{0} E_{0}}{k_{B}T}\right),
	\label{eq:thermalEscape}
\end{equation}
where $\omega_{0}$ is the normalized characteristic small amplitude
oscillation frequency of the vortex at the bottom of the potential
well.  The potential barrier height in normalized units $U_{0}(I)$ and
the attempt frequency $\omega_{0}(I)$ are calculated from
Eq.~(\ref{eq:magFieldPotential}).  $8 E_{0} = 8 \Phi_{0} j_{c} \lambda_{J}
w / (2 \pi) $ is the rest energy of the vortex, governing the energy
scale of the problem.

The activation of a vortex from a potential well can be determined
experimentally by measuring the switching current distribution of the
Josephson junction.\cite{Martinis87,Castellano96}  In this type of
measurement, the probability of the vortex escape from a pinned state
(no dc voltage drop across the junction) to a propagating state (dc
voltage drop proportional to the average vortex velocity) is
determined in dependence on the bias current.  The experimentally
observed critical current distribution $P(I)$ is to be compared with a
theoretical prediction
\begin{equation}
	P(I) = 
	\left|\frac{dI}{dt}\right|^{-1} \Gamma(I) 
	\exp\left(-\left|\frac{dI}{dt}\right|^{-1}
	\int_{0}^{I}\Gamma(I')\,dI' \right),
	\label{eq:probDistribution}
\end{equation}
which is calculated from the bias current sweep rate $dI/dt$ and the
dependence of the activation rate $\Gamma$ on the bias
current $I$.  If the escape of the vortex is due to a thermally
activated process, i.e. when $k_{b}T \geq \hbar\omega_{0}\omega_{p}$,
the critical current distribution $P(I)$ is determined by
Eq.~(\ref{eq:thermalEscape}).

\begin{figure}[bt]
\centering
\epsfig{file=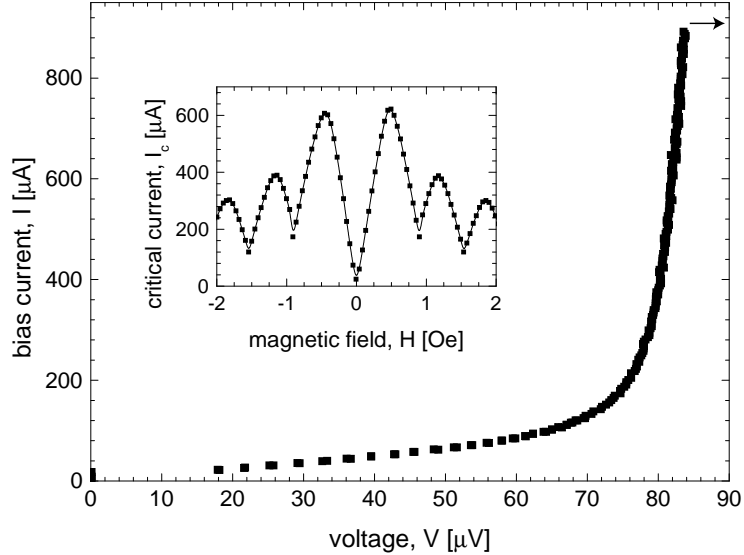, width=0.8\columnwidth}
\caption{Current-voltage characteristic of a fluxon in
the annular Josephson junction.  The modulation of critical current
$I_{c}$ with the external magnetic field $H$ is shown in the inset.}
\label{fig:IVICH}
\end{figure}
The thermally activated fluxon escape has been experimentally
investigated in a number of samples.  Here, we present data on an
annular Josephson junction of radius $r = 50 \, \rm{\mu m}$ and width
$w = 3 \, \rm{\mu m}$; its geometry is depicted in Fig.~\ref{fig:exp}. 
The junction has a critical current density of $j_{c} \approx 100 \,
\rm{A cm^{-2}}$ and, consequently, a Josephson length of $\lambda_{J}
\approx 30 \, \rm{\mu m}$.  A single Josephson vortex could be trapped
in the junction by applying a small bias current through the junction
during the transition from the normal to the superconducting state. 
The single-vortex state is identified by the lowest quantized
resonant voltage step (saturating at a about $V = 84 \, \rm{\mu V}$
for this particular sample) observed on the current-voltage
characteristics of the junction, see Fig.~\ref{fig:IVICH}.  Also, a
characteristic critical current modulation with magnetic field,
accompanied by a suppression of the critical current by a factor of
more than 100 at zero field, as shown in Fig.~\ref{fig:IVICH} and
reported earlier in Ref.~\onlinecite{Vernik97}, is observed when a
single vortex is trapped in the junction.

\begin{figure}[bt]
\centering
\epsfig{file=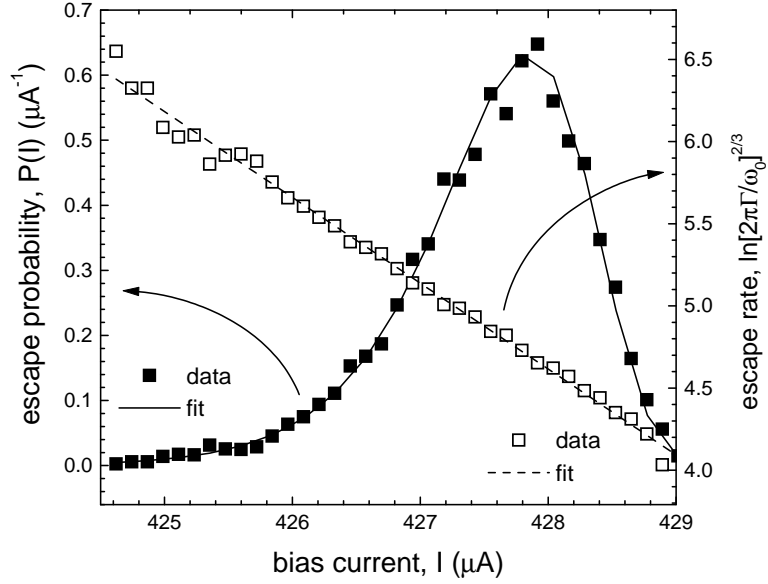, width=0.8\columnwidth}
\caption{Measured histogram (solid squares) of the critical current
distribution at $T = 6_{\rm{bath}} \, \rm{K}$ and $H = 0.25 \,
\rm{Oe}$ and calculated probability distribution $P(I)$ (solid line). 
Experimental normalized activation rate $\Gamma(I)$ (open squares) and
fit (dashed line).}
\label{fig:13PI}
\end{figure}

\begin{figure}[bt]
\centering
\epsfig{file=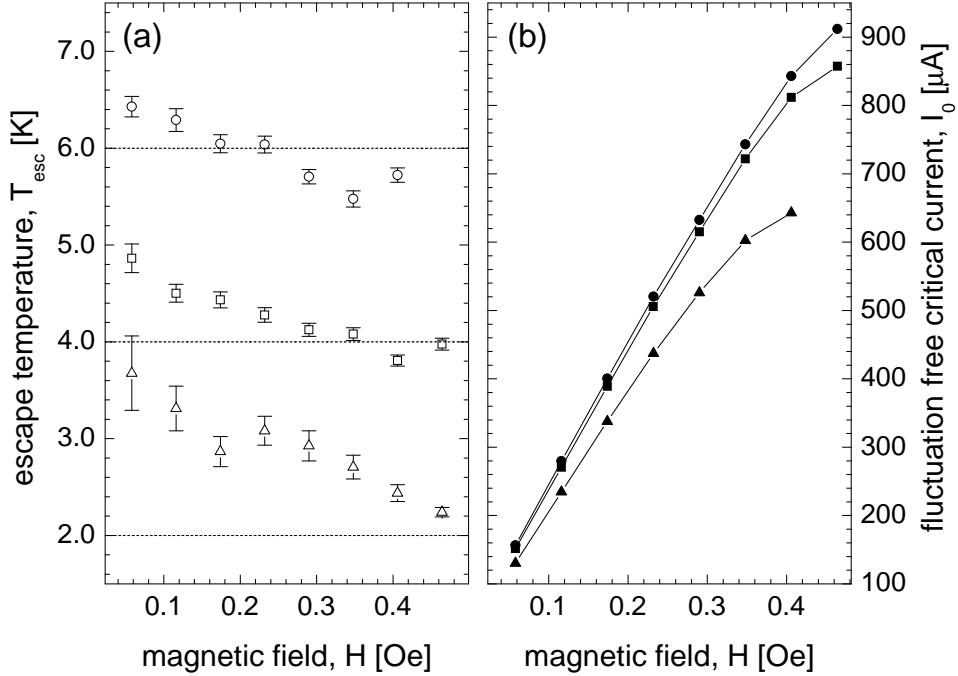, width=1.0\columnwidth}
\caption{(a) Escape temperatures $T_{\rm{esc}}$ (open symbols) and (b)
fluctuation-free critical current $I_{0}$ (closed symbols) in
dependence on the externally applied magnetic field at temperatures
$T=2$ (triangles), $4$ (squares), and $6$ (circles) K. The respective
bath temperatures are indicated by dashed horizontal lines.}
\label{fig:13Tesc}
\end{figure}

A typical measured switching current distribution $P(I)$ at $H = 0.25
\, \rm{Oe}$ and $T = 6 \, \rm{K}$ is shown in Fig.~\ref{fig:13PI}. 
The distribution is normalized such that $\int P(I) dI = 1$.  By
inverting Eq.~(\ref{eq:probDistribution}) the activation rate
$\Gamma(I)$ can be determined from the experimental data.  Using a
cubic approximation of the potential Eq.~(\ref{eq:magFieldPotential}),
the expression $(\ln{(2 \pi \Gamma / \omega_{0}))^{2/3}}$ is expected
to be linear in the bias current $I$ as indeed observed in experiment,
see data (open squares) in Fig.~\ref{fig:13PI}.
From a linear fit to $(\ln{(2 \pi \Gamma / \omega_{0}))^{2/3}}$, both
the effective temperature $T_{\rm{esc}}$ of the thermal escape and the
critical current in the absence of fluctuations $I_{0}$ can be
determined self-consistently.  The results of this analysis are shown
in Figs.~\ref{fig:13Tesc}a and b for the bath temperatures
$T_{\rm{bath}} = 2,\, 4,\, 6 \, K$ and magnetic fields in the range of
$H = 0 \ldots 0.5 \, \rm{Oe}$.  The error bars in
Fig.~\ref{fig:13Tesc}a indicate the uncertainty in the value of
$T_{esc}$ calculated using the fitting procedure explained above.

At high temperatures, the experimentally determined values of
$T_{\rm{esc}}$ are within $1 \, K$ of the thermal bath temperatures
$T_{\rm{bath}}$ for all values of the magnetic field.  For the lowest
temperature, the fitted escape temperature is notably higher than the
bath temperature, which is due to the limited current resolution of
approximately $0.5 \, \rm{\mu A}$ in the $1 \, \rm{mA}$ range of the
measurement setup used for data acquisition in this particular
experiment. Moreover, a small but systematic increase of the escape
temperature for small magnetic fields is observed.  This effect may be
explained by considering a localized and magnetic field independent
contribution to the pinning potential\cite{Fistul00}, which leads to
an enhanced activation rate at low fields.  In Fig.~\ref{fig:13Tesc}b,
the magnetic field dependence of the fluctuation-free critical
currents $I_{0}(H)$ is plotted.  The currents are observed to be
larger than the critical currents measured in the $I_{c}(H)$ patterns,
as expected.

\section{QUANTUM TUNNELING}
In the limit of $T \rightarrow 0$, the thermal activation of the vortex
from the potential well is exponentially suppressed and eventually the
temperature independent contribution due to quantum tunneling of the
vortex through the potential barrier will be the dominating escape
process.  By measuring the switching current distribution we expect to
observe a crossover from thermal activation to quantum escape at low
temperatures.  Experiments aiming to observe this effect are currently
in preparation.

The theory of macroscopic quantum tunneling for a single Josephson
vortex has been recently developed and discussed in detail in
Refs.~\onlinecite{Kato96,Shnirman97b}.  Essentially, the quantum
tunneling of the vortex can be described in the Caldeira- Leggett
approach\cite{Calderia81}, where the escape rate is given by
\begin{equation}
	\Gamma_{\rm{qu}} = A \exp(-B)
	\label{eq:WKBrate}
\end{equation}
with
\begin{eqnarray}
	A & = & \sqrt{60} 
	\omega_{0}\omega_{p}\left(\frac{B}{2\pi}\right)^{1/2} (1+O(\alpha)),
	\label{eq:coeffA}  \\
	B & = & \frac{36 U_{0} E_{0}}{5 \hbar \omega_{0}\omega_{p}} 
	(1+1.74\alpha+O(\alpha^{2})).
	\label{eq:coeffB}
\end{eqnarray}
This analysis is valid for an expansion of the potential $U(q)$ up to
third order in the vortex coordinate $q$.  Our numerical and
analytical calculations of the expected switching current
distributions (\ref{eq:probDistribution}) using
Eqs.~(\ref{eq:thermalEscape},\ref{eq:WKBrate}) show that a cross-over
from thermal activation to quantum tunneling can be expected at
temperatures $T \approx 20 \, \rm{mK}$ for junctions which are several
micron wide.  
In these calculations we took into account the effects of
dissipation which can be modeled using the effective damping parameter
$\alpha$ in Eq.~(\ref{eq:coeffB}).  We considered both the
quasiparticle tunneling across the junction barrier and the
quasiparticle impedance of the electrodes as possible origins of
dissipation (also see Ref.~\onlinecite{Kato96}).  At $4.2 \, \rm{K}$
the damping parameter $\alpha$ is typically $10^{-2}$ and decreases
exponentially with temperature.  Therefore, this damping mechanism
does not substantially decrease the cross-over temperature.  In
Ref.~\onlinecite{Shnirman97b}, the dissipation induced by the interaction of
the vortex with plasma waves is studied, and the resulting decrease of
the tunneling rate is found to be small.  So far no other interaction
mechanisms of the system with the environment have been taken into
account.

\begin{figure}[tbp]
\centering
\epsfig{file=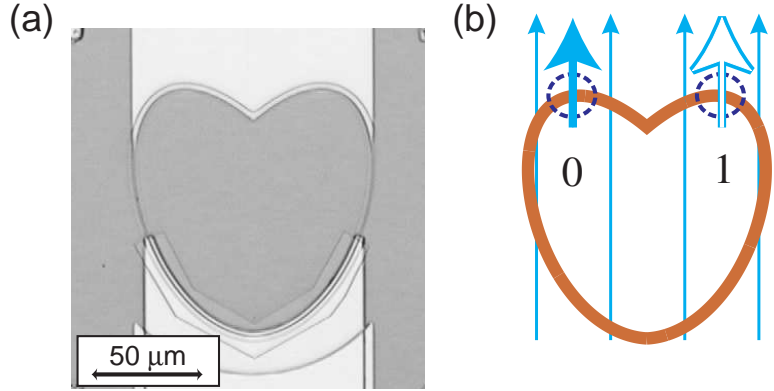, width=0.8\columnwidth}
\caption{(a) Photograph of a $0.3 \, \rm{\mu m}$ wide heart-shaped
junction fabricated in Nb-AlO$_{x}$-Nb technology.  (b) Sketch of
heart-shaped junction with magnetic field applied along the symmetry
axis of the junction.  Two energetically identical vortex states are
indicated by arrows in positions $0$ and $1$.}
\label{fig:heart}
\end{figure}

Our calculations show that the cross-over temperature can be
increased close to $100 \, \rm{mK}$, by reducing the junction width down
to sub-micron size and optimizing the junction electrical properties.
Most experiments on fluxon dynamics conducted so far have been done
using long Josephson junctions of width $w > 3 \rm{\mu m}$.  This
limit is imposed by the standard photolithographic preparation
procedure.  Recently, we succeeded to fabricate high quality
ultra-narrow long Josephson junctions based on niobium-aluminum-oxide
trilayer technology of width down to less than $0.3\,\rm{\mu
m}$.\cite{Koval99}  In this procedure, we are using electron-beam
lithography for junction definition and cross-linked PMMA for junction
insulation.\cite{Koval99}  We have successfully prepared and
characterized narrow long junctions of different geometries.  A
photograph of a $0.3 \, \rm{\mu m}$ wide heart-shaped long Josephson
junction is shown in Fig.~\ref{fig:heart}a.  First experiments on the
activation of fluxons from magnetic field-induced potentials in $0.3
\, \rm{\mu m}$ wide annular junctions have been recently performed and
will be presented elsewhere.

\section{POTENTIAL ENGINEERING}
\begin{figure}[tbp]
\centering
\epsfig{file=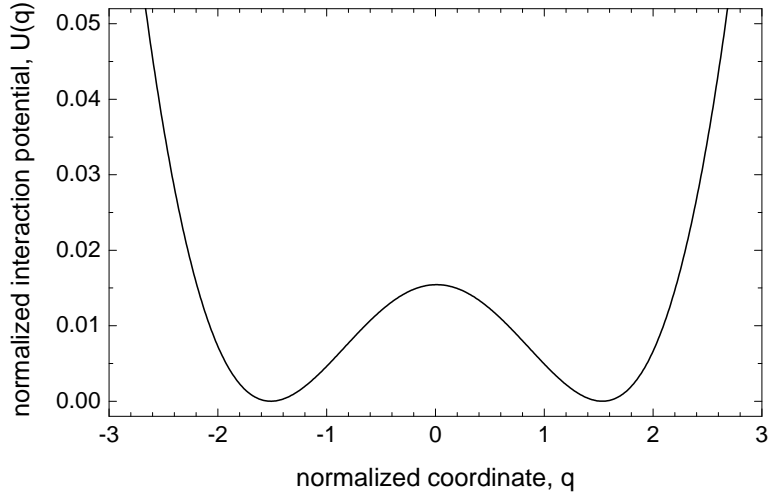, width=0.8\columnwidth}
\caption{Normalized magnetic field interaction potential $U(q)$ at $h
= 1$ calculated according to Eq.~(\ref{eq:magFieldPotential}).}
\label{fig:pot}
\end{figure}

Since the interaction energy of the vortex with the external magnetic
field is determined by the shape of the junction according to
(\ref{eq:magFieldPotential}), it is possible to design an experiment
with, in principle, arbitrarily shaped fluxon potential by simply
varying the junction geometry.  Since the major goal in our
experiments with quantum fluxons will be a demonstration of
macroscopic quantum coherence, one needs to realize a double-well
potential for a vortex in the junction.  We suggest to use a
heart-shaped Josephson junction, as shown in Fig.~\ref{fig:heart}a. 
The magnetic field $\vec{H}$ applied along the symmetry axis in the
junction plane forms two equal potential wells for the fluxon at
locations "0" and "1".  At these two positions the fluxon magnetic
moment $\vec{\mu}$ is directed along $\vec{H}$, see
Fig~\ref{fig:heart}b.  The barrier height between the wells can be
conveniently controlled by the field amplitude and its angle with the
symmetry axis of the junction.  The corresponding potential profile at
$h= 1$, calculated from Eq.~(\ref{eq:magFieldPotential}) with the
normalized junction length $L/\lambda_{J} \approx 35$, is plotted in
Fig.~\ref{fig:pot} in units of $E_{0}$.  We note that there are
constraints on the potentials that can be generated by changing the
junction shape, because the shape dependent part of the potential is
effectively averaged over the vortex size ($\propto \lambda_{J}$), as
can be seen from Eq.~(\ref{eq:magFieldPotential}).  In the limit of
$\lambda_{J} \rightarrow \infty$ all features of the potential
disappear.

In the quantum limit, the two distinct fluxon states in the double
well potential may be employed as a degenerate two-state system. 
Under sufficient decoupling from the environment and provided that the
temperature and dissipation in the junction are low enough, the
superposition of the macroscopically distinct quantum states "0" and
"1" is expected to be observed.  The preparation of the initial state
for a quantum coherence measurement may be achieved by turning the
magnetic field in the plane of the junction, and the final state of
the vortex can be read out by performing an escape measurement from
one of the potential wells.

\section{CONCLUSIONS}
We have considered a vortex in a long Josephson junction interacting
with an externally applied magnetic field.  Experiments indicating the
thermal activation of the vortex from the field-induced potential have
been successfully performed.  Moreover, the prospects of observing
quantum tunneling of a single vortex from such a potential are
discussed as well as possibilities of potential engineering in order
to form a double-well potential for use in a qubit.

\section*{ACKNOWLEDGEMENTS}
This work was supported by the Deutsche Forschungsgemeinschaft (DFG). 
We would like to thank M.~G.~Castellano, M.~Cirillo, M.~Feldman,
A.~Kemp, Yu.~Makhlin, B.~A.~Malomed, J.~E.~Mooij, P.~M{\"u}ller,
G.~Sch{\"o}n, K.~Urlichs, and C.~H. van der Wal for useful discussions.

\nopagebreak

\end{document}